\documentclass[reprint,twocolumn,preprintnumbers,pre,amsmath,amssymb]{revtex4}
\usepackage{dcolumn}% Align table columns on decimal point
\usepackage{hyperref}% add hypertext capabilities
\usepackage{color}
\usepackage{graphicx}
\usepackage{bm}
\usepackage[justification=raggedright]{caption}
\usepackage{amsmath}
\begin{document}

\title{Temporal complexity in emission from Anderson localized lasers }
\author{Randhir Kumar}
\author{M. Balasubrahmaniyam}
\author{K Shadak Alee}
\author{Sushil Mujumdar}
\email[]{mujumdar@tifr.res.in}
\homepage[]{http://www.tifr.res.in/~mujumdar}
\affiliation{Nano-optics and Mesoscopic Optics Laboratory, Tata Institute of Fundamental Research, 1, Homi Bhabha Road, Mumbai, 400 005, India}
\date{\today}
\begin{abstract}
Anderson localization lasers exploit resonant cavities formed due to structural disorder. The inherent randomness in the structure of these cavities realizes a probability distribution in all cavity parameters such as quality factors, mode volumes, mode structures etc, implying resultant statistical fluctuations in the temporal behavior. Here, we provide the first, direct experimental measurements of temporal width distributions of Anderson localization lasing pulses in intrinsically and extrinsically disordered coupled-microresonator arrays. We first illustrate signature exponential decays in the spatial intensity distributions of the lasing modes that quantify their localized character, and then measure the temporal width distributions of the pulsed emission over several configurations. We observe a hitherto-unreported dependence of temporal widths on the disorder strength, wherein the widths show a single-peaked, left-skewed distribution in extrinsic disorder and a dual-peaked distribution in intrinsic disorder. We propose a model based on coupled rate equations for an emitter and an Anderson cavity with a random mode structure, which gives excellent quantitative and qualitative agreement with the experimental observations. The experimental and theoretical analyses bring to the fore the temporal complexity in Anderson localization based lasing systems.
\end{abstract}
\maketitle

The common belief that structural imperfections are detrimental to lasing action has been challenged by the concept of random lasing\cite{wiersmareview,caoreview,letokhov68,lawandy94,pradhan94,cao99,genack_loc05,cao00,mujumdar04prl,tiwari_prl,uppu_prl,liu14,conti08,leonetti11,ghofrania15,stano13}.
This phenomenon thrives on structural randomness, wherein scattering-induced feedback realizes lasing in the presence of gain. Within the substantial body of literature on random lasers, the most contemporary reports deal with the studies of lasing in the Anderson localization regime. Anderson localization realizes disorder-induced trapping of waves\cite{anderson58}, and a large experimental effort has been focused on the study of light localization\cite{segev13,billy08,john87,wiersma97,deraedt89,schwartz07,lahini08,mookherjea14,sapienza10,nahata, karbasi12,szameit12,topolancik07,caze13,leseur14,riboli11,leonetti14prl,szameit10,vicencio15}. Importantly, the fact that Anderson-localized states are resonant states is obviously consequential in lasing, as they can provide the requisite cavities for efficient lasing. In a set of early pioneering investigations in semiconductor powders\cite{cao99,cao00}, resonant random lasing involving recurrent scattering was demonstrated. Such recurrent feedback, one of the key ingredients for Anderson localization, was inferred from the unique spectral signatures of the resonant modes. Apart from spectral properties, temporal dynamics of resonant random lasing was also reported, wherein the signatures of the cavities were observed in relaxation oscillations\cite{soukoulis02}. Interestingly, experiments also reported relaxation oscillations in nonresonant random lasers\cite{molen09,noginov04}. These experimental studies enlisted the temporal behavior of generic random lasers outside the Anderson-localized regime.

In three-dimensional systems, resonant states are rarely Anderson-localized, due to a stringent requirement of critical disorder\cite{lee85}. Also, under appropriate conditions of gain, the spectral signatures of resonant modes may also be emulated by nonresonant systems\cite{mujumdar04prl,mujumdar07}. In contrast, all eigenstates of lower dimensional structures are necessarily localized, sample size permitting. Specifically, the crossover into Anderson localization happens when $\xi<L$, where $\xi$ characterizes the localization length. Several theoretical studies of one-dimensional systems have revealed a substantial physics of localization\cite{lifshitsbook,gredeskul,bliokhbook,rybin}. This has motivated a large experimental effort in lower-dimensional samples\cite{karbasi12,lahini08,schwartz07,leonetti14prl,szameit12,szameit10,vicencio15,nahata,mookherjea14,sapienza10,riboli11}. Perhaps the most important achievement in low-dimensional disordered systems is the direct measurement of the exponentially decaying photon wavefunction\cite{billy08,lahini08,schwartz07,nahata}. This exponentially decaying localized wavefunction is the classic signature of Anderson localization\cite{anderson58}, and provides an unambiguous evidence thereof. In the lasing domain, too, there have been significant studies in lower dimensions. A one-dimensional dielectric multilayer with a gain medium\cite{genack_loc05} exhibited spectral signatures of photon localization lasing. A fiber-based one-dimensional random laser was used to illustrate the transmission spectra of localization-based cavities\cite{bliokh12}. Another system exploited disorder-induced localized modes in photonic crystal waveguides to realize ultrasmall sized nanolasers\cite{liu14}. These studies provided significant information on localized lasing modes using their spectral signatures. To our knowledge, no temporal studies exist in the domain of Anderson-localized random lasers. In contrast to conventional cavities, Anderson cavities exhibit a statistical variation of spatial profiles, quality factors and mode volumes\cite{liu14}. Furthermore, while conventional cavities are designed to have a homogeneous mode profile, Anderson cavities, by definition, have a random mode structure. Due to these factors, the temporal behavior of Anderson-localized lasing modes is not intuitively obvious and needs to be experimentally documented. Furthermore, these experiments need to involve multiple Anderson-localizing configurations in order to obtain statistically consistent datasets. In this paper, we report the temporal complexity of the Anderson localized laser through the measurement of the temporal width distributions of the laser pulses. We employ a one-dimensional disordered system constituting a linear array of coupled microresonators. The array acts as a crystal in which the states at the bandedges have been shown to localize under disorder\cite{mookherjea14}. Our system offers fine control of disorder strength in transiting from a nearly-periodic configuration sustaining only intrinsic disorder to intentionally disordered configurations. After endorsing the localization via direct measurements of exponentially decaying wavefunctions, we measure the pulsewidth distributions of the modes. We observe that the intentionally disordered system exhibits a left-skewed single-peaked distribution of temporal widths, whereas the intrinsically disordered system reveals a bimodal distribution. We model the system using coupled emitter-cavity rate equations which are suitably modified to invoke the random mode profiles of the Anderson cavity. The model provides excellent agreement with the experimental results.

The experiments were performed on a coupled microresonator array made out of Rhodamine-6G dye dissolved in alcohol. A one-dimensional `crystal' was obtained by realizing a periodic array of monodisperse microresonators using a vibrating orifice aerosol generator, which is based on perturbation of unstable liquid jets\cite{lin90}. When a liquid is forced through a narrow orifice, usually under pressure of a non-interacting gas, it emits as a cylindrical liquid jet that is mechanically unstable. Under periodic perturbation with the appropriate amplitude and frequency, the jet breaks up into equal-sized microdroplets, the radius of which is given by $a=(3F/4 \pi f)^{1/3}$, where $F$ is the liquid flow rate and $f$ is the frequency of the vibrating orifice\cite{berglund73}. Physical parameters such as viscosity of the liquid and orifice diameter etc are implicit in  the flow rate. In our experiments, the orifice diameter of a microcapillary was 10~$\mu$m. The output end of the microcapillary was fit with a piezo-activated gate that induced periodic perturbations as required. For parameters such as in the intervals of $F$ = 60-120~$\mu$l/min, and $f$ = 600-100~kHz, we obtained microdroplet diameters in the neighborhood of 16-20$~\mu$m. These parameters are corroborated by the diameter measurements using separation in Mie resonant modes, which is a more accurate method. Importantly, monodisperse and periodic arrays are formed for the correct combinations of flow rate and piezo-frequency. So, extrinsic disorder could be obtained simply by deviating from the prescribed frequency. Furthermore, even under `perfect periodic' conditions, the inherent fluctuations in the fluid dynamic process results in diameter fluctuations, which constitute the intrinsic disorder. The solvent used in our experiments was 1:1 measure of methanol and ethylene glycol, so as to obtain sufficient viscosity to minimise size fluctuations and still maintain a high gain coefficient.

\begin{figure}
\includegraphics[width=8.5cm]{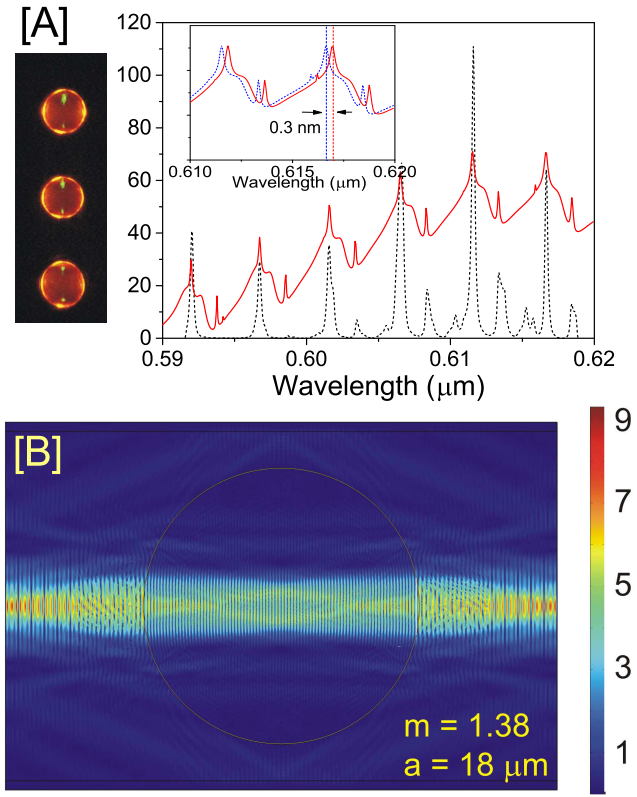}
%\vskip 2cm
\caption{\label{fig:setup2}  [A] Sizing of the microresonators (image on left), using the whispering gallery modes. Black dotted spectrum is the experimentally observed spectrum of the individual microresonators. The red solid spectrum shows the Mie theoretic calculation giving a perfect match for the peak positions, for a diameter of 18.082~$\mu$m. Inset shows two computed spectra for diameters that are different by 10 nm realizing peak shift of 0.3 nm. [B]  Field distribution in a unit cell of the microresonator array, as computed using the finite element method in two dimensions applying periodic boundary conditions in the horizontal direction.}
\end{figure}

The image in Fig~1[A] depicts three spherical microdroplets in the array. A spherical optical resonator sustains whispering gallery modes that are very high quality resonances, which allow us to quantify the disorder. For large microspheres with refractive index $m$, the diameter $a$ and the central wavelength $\lambda_o$ of the WGMs' are related as $ a = \frac{\lambda_o^2}{\pi \Delta\lambda}\frac{tan^{-1}\sqrt{(m^2-1)}} {\sqrt{(m^2-1)}}$\cite{campillotextbook}, where $\Delta\lambda$ is the free spectral range. Accordingly, a differential change of $\sim3$~nm in the diameter causes a differential of $\sim0.1$~nm in $\lambda_o$, as shown in Fig 1(a), which is easily measured in our spectrometer, which has an instrument response of $0.1$~nm. For the results presented in this work, we realized microresonator diameter and spacing of 18~$\mu$m each. When the aerosol generator was run at the best monodispersity, the inherent diameter fluctuations were measured to be about $\pm5$~nm ($<0.4$~nm in resonant wavelength). This describes the  intrinsic disorder in our crystalline array. When the frequency of the piezogate at the output of the microcapillary was deliberately tweaked, larger diameter fluctuations were achieved, realizing extrinsic disorder in the array. The extrinsically disordered system had diameter fluctuations of $\pm50$~nm which sufficed to show a significant difference in the temporal behavior.

While these whispering gallery modes are the natural highest quality resonances of an individual microsphere, the mode excited under periodic boundary conditions in one direction is a unidimensional axial mode, as clarified in Fig 1[B]. The figure depicts a two-dimensional finite-element computation of a mode of the resonator (diameter 18~$\mu$m, refractive index 1.38) under periodic boundary conditions when the periodicity is applied only in longitudinal direction (here, Z direction). To maintain linearity within a mesh element, the mesh element is kept below $\lambda/8m$ in the dielectric region, where $m$ is the refractive index of the dielectric medium. Floquet periodic boundary condition was maintained along the longitudinal direction for periodic boundary, and perfectly matched layers for open boundaries in the transverse direction. Eigenfrequencies and corresponding field distributions were calculated. It can be seen that the field intersects the dielectric-air interface over a very small arc, due to the tight axial nature of the mode. The transverse extent of the field is $\sim3.5~\mu$m. Furthermore, the phasefronts are nearly flat, because $a >> \lambda/m$. This motivates the use of a plane parallel resonator to approximate the curved resonator. There is an added motivation for this approximation. One-dimensional transport can be accurately calculated using the transfer matrices technique, which is computationally inexpensive\cite{jiang02,andreasen11}. This enables the calculation for sufficiently large number of configurations to obtain configurationally-averaged parameters, which are the quantities of interest in the physics of disorder. Subsequent computations of the microresonator array were hence carried out using transfer matrix computations.

\begin{figure}
\includegraphics[width=8.5cm]{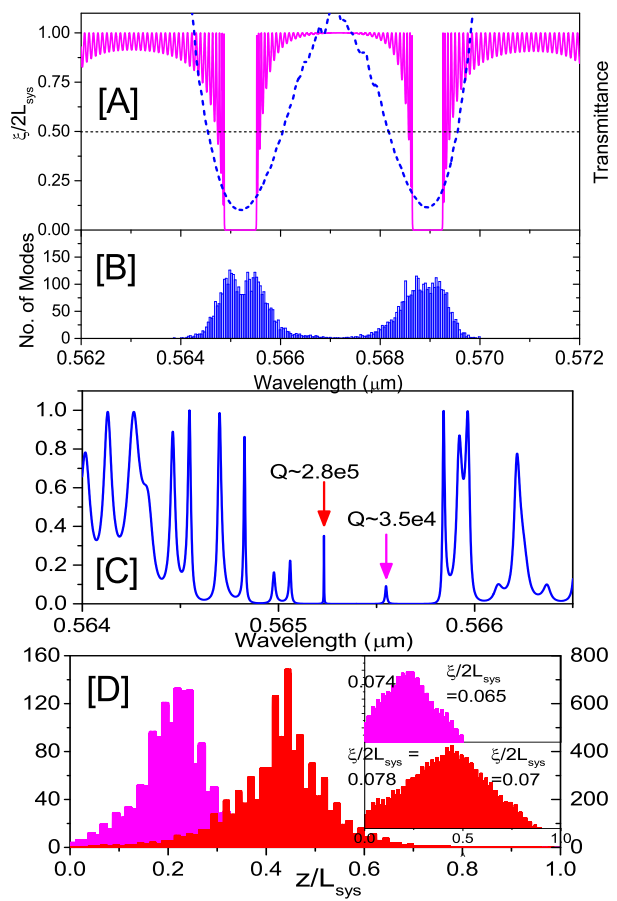}
\caption{Computed localization behavior for our system. [A] The magenta plot (solid line) marks the transmission spectrum of a periodic dielectric-air (18~$\mu$m thick each) binary multilayer. The dotted blue curve depicts the half localization length $\xi/2$ as a function of wavelength, for extrinsic disorder. [B]: Histogram illustrating the probability of occurrence of Anderson localized lasing modes in the  system. [C]: Transmission for a particular configuration, of which two modes are marked by colored arrows. [D]: The spatial profiles of the two marked modes. Insets show the same in log scale, quantifying the $\xi/2L_{sys}$. }
\end{figure}

Figure 2 depicts known spatio-temporal properties of Anderson-localized modes in a one-dimensional randomized crystalline system, computed using transfer matrices for our sample parameters. The transmission (Fig~2[A], solid magenta curve) is zero in the bandgap region in the crystal. Introduced  disorder realizes edge states and gap states, which lase in the presence of gain. The dotted blue curve depicts the configurationally-averaged half localization length $\langle\xi/2\rangle$ $= -1/\langle \ln T \rangle$  for extrinsic disorder. The denominator 2 in $\xi/2$ relates to the intensity $|\Psi(z)|^2$ which is measurable, and not the field $\Psi(z)$. All linear dimensions are normalized to the system size $L_{sys}$ in this report. The histogram in [B] shows the probability of occurrence of the lasing modes on the wavelength axis. Clearly, the $\xi/2 < 0.5$  in the spectral region of the maximum probability of occurrence, implying that these lasing modes will possess an Anderson character. Figure 2[C] shows the spectral signature (blue curve) of one random manifestation, of which two modes are identified in red (left arrow) and magenta (right arrow) arrows. Spatial distribution of the two modes (Fig 2[D]) shows the exponentially decaying wings induced by Anderson localization. The insets show the modes on a  logarithmic scale, yielding the $\xi/2 \sim0.07$. The two modes have very comparable spatial localization, but significantly different Q factors. The red mode (centrally located, scaled on right Y-axis) has a Q-factor about an order of magnitude larger than the magenta mode (scaled on left Y-axis), on virtue of its central location in the system. Besides, the spatial overlap between the modes is conducive to gain competition between the modes. Evidently, the temporal dynamics of the two modes under lasing conditions will be significantly different. The experiments in this work were designed to capture the resulting temporal behavior of Anderson modes.

\begin{figure}
\includegraphics[width=8.5cm]{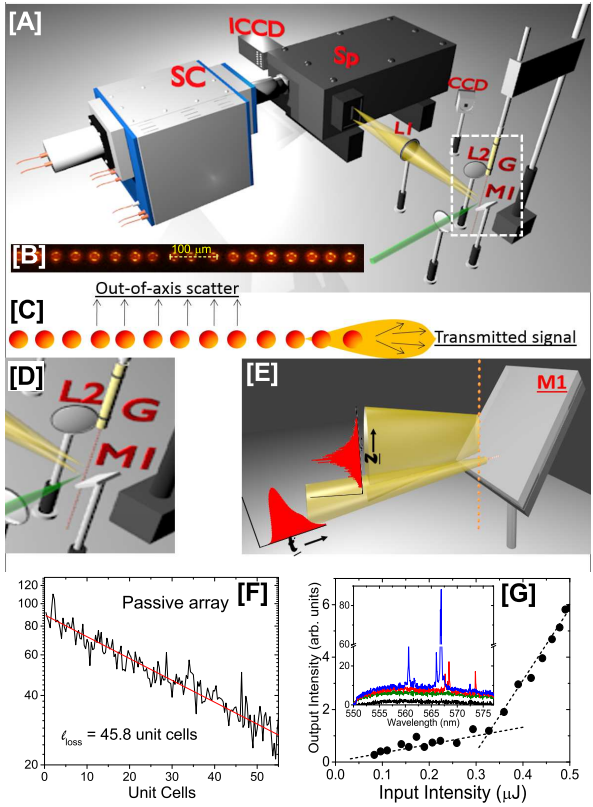}
%\vskip 2cm
\caption{\label{fig:setup2} [A] Schematic of the experimental setup, with the array generator `G' enclosed in the white dashed rectangle. See text for experimental details. [B]The microresonator array.  [C] Schematic of the angular profile of the transmitted light. Also marked is the direction of out-of-axis scatter. [D]: Emphasized image in the white rectangle in [A]. [E] emphasizes the role of mirror M1, which reflects the on-axis transmission carrying temporal information into the Spectrometer. The out-of-axis scatter carried the spatial information.[F] Loss characterization for the passive array. [G] Lasing threshold measurement (output intensity as a function of input) under optical pumping. Inset: Spectral narrowing with increased pump energy.}
\end{figure}

Figure 3[A] shows the schematic of the experimental setup, while [B] shows the image of the coupled microresonator array. The white dotted rectangle in [A] demarcates the generator ($G$) of the coupled microresonator array, and the illumination of the array. Under optical excitation provided by a frequency-doubled Nd:YAG laser (pulse-width $\sim60$~ps, rep rate 10 Hz, shown as the green beam), the emission propagating along the array axis is transported according to the photonic bands of the array. Fig 3[C] shows a schematic of the array emission, illustrating the plume of the transmitted signal, and its out-of-axis scatter in the transverse direction. [D] shows the magnified image of the dotted rectangle, emphasising the generator $G$, the emitted microresonator array, and the two emitted beams (signal and out-of-axis scatter) shown in yellow. Fig 3[E] emphasizes the role of mirror $M1$, which reflects the transmitted signal into the dual-output Spectrometer $Sp$. The angular spread of the transmitted plume allows for reflection of the signal without interfering with the falling array. This carried the temporal signature of the mode. The out-of-axis scattered light carries the information of the spatial distribution of the mode, and is directly imaged onto the input slit of the spectrometer.   One output of the spectrometer was connected to the Streak Camera $SC$, while the other one was attached to an ICCD. A flippable mirror at the output port of the spectrometer allowed to switch between the two devices. The spatial profiles of the modes were spectrally resolved and imaged directly onto the ICCD. The out-of-axis light constitutes the scattering loss in the passive system. This was measured in separate experiments on the passive array, and the loss was characterized by the loss length to be $\sim46$ unit cells (Fig 3[F]), where one unit cell comprises one microresonator plus one spacing. For the lasing experiments, the crystal length was chosen to comprise 40 unit cells, less than the loss length. Importantly, the systemic losses are overcome by the gain introduced by the optical pumping. Fig 3[G] characterizes the system under excitation, i.e, in the lasing regime. The inset illustrates four typical spectra at pump energies of $\sim0.19~\mu$J (black curve), $\sim0.26~\mu$J (green curve), $\sim0.32~\mu$J (red curve) and $\sim0.36~\mu$J (blue curve).  The main plot shows the integrated intensity in the band $555~$nm to $575~$nm as a function of input pump intensity, elucidating a clear threshold at $\sim0.32~\mu$J. This behavior is very similar to the earlier reported localization laser\cite{genack_loc05} when the localization based cavities were shown to lase. At the threshold pumping, the system becomes conservative\cite{liu14}. Our experiments were run just above the threshold pumping.

Under conditions of randomness, the states that lase are identified to be the gap states that are created in the bandgaps of the erstwhile crystal\cite{tiwari_prl}. As shown in Fig~2, the localization length minimizes in the same region for the randomized crystal.  This implies that the observed lasing states will be Anderson-localized, provided the crystal size is large enough.  Accordingly, we measured the existence and extent of localization by directly imaging the spatial eigenfunction of the modes. Indeed, the eigenfunctions exhibited clear exponential tails as expected from Anderson localization. Figure 4[A] and [B] illustrate two modes, co-existing in a single configuration, spatially resolved by the ICCD. The modes are peaked at random positions in the vicinity of the center of the sample. On either side of the peak, the modes clearly exhibit a tight exponential decay in the wings. The insets show the same modes on a logarithmic scale. In general, the decay length is determined by $[\ell_{loss}^{-1} + \xi^{-1}]^{-1}$. With the gain at or just above threshold, the system is conservative and $\ell_{loss} \rightarrow \infty$. The decay arises purely from Anderson localization and the decay length has been used to characterize the localization length\cite{billy08}. The red mode shows an overall symmetric character with $\xi/2 \sim 0.16$, and peaks at $z \sim 0.46$. The magenta mode is asymmetric, with $\xi/2 \sim 0.17$ in the left wing and $\xi/2 \sim 0.09$ in the right wing, with the mode peaking at $z \sim 0.65$. The distribution of $\xi/2$, measured over 300 individual modes, (Fig~4[C]) shows an asymmetric distribution about the peak at $\xi/2 \sim 0.185$. The mean localization length $\langle \xi/2 \rangle$ from this distribution is 0.2. In the presence of multiple configurations, Anderson localization is identified and $\xi$ is determined from the ensemble-averaged mode profile\cite{schwartz07,lahini08,segev13}. Figure 4[D] exhibits the ensemble-averaged mode profile, showing an almost symmetric exponential decay on either side, yielding $\xi/2 \sim 0.185 \pm 0.005$. These measurements endorse the tightly localized character of the Anderson modes in our system.

\begin{figure}
\includegraphics[width=8.5cm]{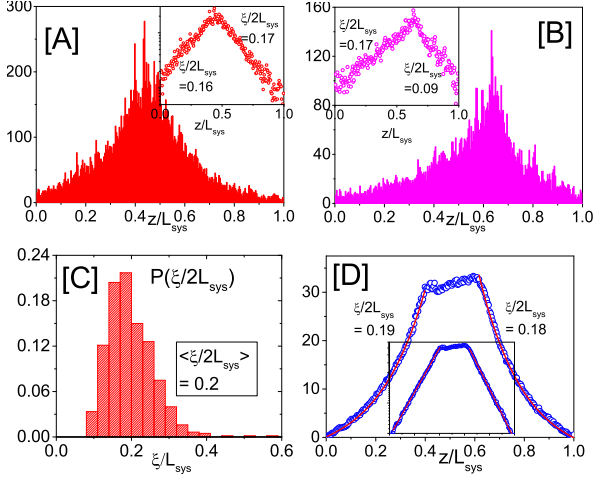}
\caption{\label{fig:setup2} [A] \& [B] Spatial intensity distributions of two modes in a single configuration showing exponentially decaying wings. Insets: Mode profiles on logarithmic axes, characterising the $\xi/2$. [C] Distribution $P(\xi/2)$ measured over individual modes. [D] Ensemble-averaged mode profile exhibiting the averaged exponential profile (inset on log scale) of Anderson localization.}
\end{figure}

Once Anderson localization was endorsed, temporal measurements were carried out using the Streak Camera. Figure 5[A] shows the temporal profiles of all lasing modes in one configuration under extrinsic disorder. The system consistently offered multimode emission, with considerable heterogeneity in parameters such as startup time, risetime, peak intensity, temporal shapes and decay time. Of particular interest is the statistical behavior of the temporal width $\Delta\tau$ (the full-width-at-half-maximum), which characterizes the pulsewidth. The distribution $P(\Delta\tau$) over a thousand pulses (Figure 5[B]) peaks at 65~ps, and exhibits a left-skewed behavior. Thus, the $P(\Delta\tau < 65$~ps) steadily decreases, while it falls rapidly on the other side of the maximum. Increasing the disorder strength did not significantly alter the shape of the distribution. On the contrary, lowering the disorder to intrinsic disorder led to a striking behavior. Figure 5[C] depicts the temporal output for intrinsic disorder, showing five simultaneously excited modes. The modes show similar heterogeneity in various parameters as in Fig 5[A]. A prominent multi-peaked character is also seen in some modes. Unlike the situation of extrinsic disorder, the distribution $P(\Delta\tau)$ (Fig 5[D]) shows a bimodal character. In addition to the peak $\sim 64$~ps, a second peak was manifested at around $\sim 36$~ps, suggesting the generation of newer modes with \emph{lower} temporal widths.

\begin{figure}
\includegraphics[width=9.5cm]{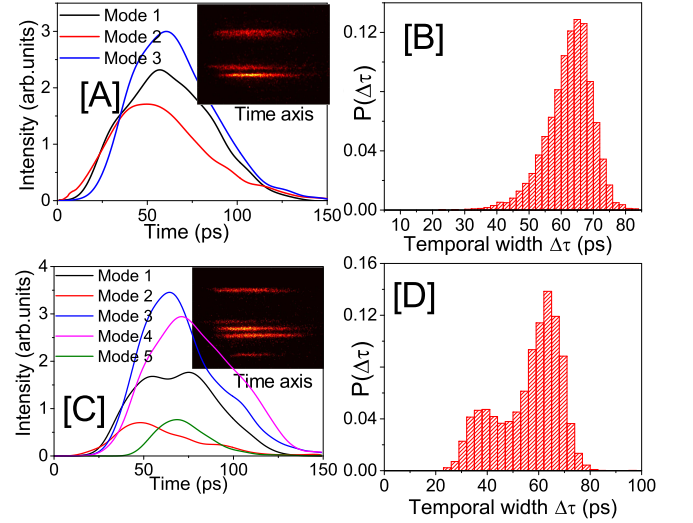}
\caption{\label{fig:figure3} [A] Three simultaneously excited lasing modes in a particular configuration of disorder. Inset: Streak image of the modes, numbered vertically downwards. [B] Experimentally measured distribution of temporal widths P($\Delta \tau$) of the lasing modes, showing a left-skewed distribution. [C] Temporal behavior under intrinsic disorder. Five lasing modes in one configuration are depicted. [D] The corresponding P($\Delta \tau$), showing a unique bimodal distribution. }
\end{figure}

To understand the overall temporal behavior of the Anderson modes, we set up a theoretical model based on coupled emitter-cavity rate equations \cite{siegman_book}. A similar scheme has been used in random lasers earlier in the diffusion regime\cite{noginov04,molen09}, wherein resonant cavities do not exist. In our localizing samples, we have explicit cavities with an observable mode structure. Therefore, our model was modified such that the cavity had a random mode profile that cannot be described analytically. To that end, we divided the one-dimensional cavity into several slabs indexed by $j$. Accordingly, the coupled rate equations are

\begin{eqnarray}
  \frac{dn_{ij}}{dt} &=& K_{ij}N_{2,j}(n_{ij}+1)-\gamma_{c,i} n_{ij} \\
  \frac{dN_{2,j}}{dt} &=& R_p -\displaystyle\sum_{i} K_{ij} N_{2,j} n_{ij} -\gamma_2 N_{2,j}
 \end{eqnarray}

Here, Eq.~1 describes the evolution of the photon number $n_{ij}$ in the $i^{th}$ mode at the position $j$ of the cavity, with $\gamma_{c,i}$ being the cavity decay rate of the $i^{th}$ mode. Eq.~2 tracks the upper level population $N_2$ at the position $j$, for a given pumping rate $R_p$ and spontaneous decay rate $\gamma_2$ to lower levels. The two equations are coupled through the coefficient $K_{ij}$ which, for a lorentzian transition, reads as
\begin{equation}
K_{ij}=\frac{3}{4\pi^2} \frac{\omega_{em} \gamma_{rad} \lambda_i^3}{\Delta\omega_{em} V_{c,ij}};
\end{equation}
where, $\omega_{em}$ and $\Delta\omega_{em}$ indicate the central frequency and linewidth of the emitter transition and $\gamma_{rad}$ indicates the radiative rate over the lasing transition. The mode structure was invoked by defining the effective cavity mode volume experienced by the emitter at longitudinal position $j_0$ in the $i^{th}$ mode as $V_{c,ij_0}=\frac{\int_j{\epsilon_{j} E_{ij}^2}{dx^3}}{(\epsilon_{j_0} E_{ij_0}^2)}$\cite{maier_opex}. Figure 6 exhibits the consequence of this modification. Panel [A] shows the actual mode profile of one particular passive mode. The panel [B] displays the effective mode volume experienced by a molecule as a function of its position. Thus, the molecule placed at the maximum of the mode experiences minimum mode volume. Inset in [B] emphasizes the region of very small mode volume. The panel [C] displays the consequent coupling coefficient. Clearly, the technique incorporated the random spatial structure of the Anderson cavities, which sets them apart from the conventional laser cavities.

\begin{figure}
\includegraphics[width=8.5cm]{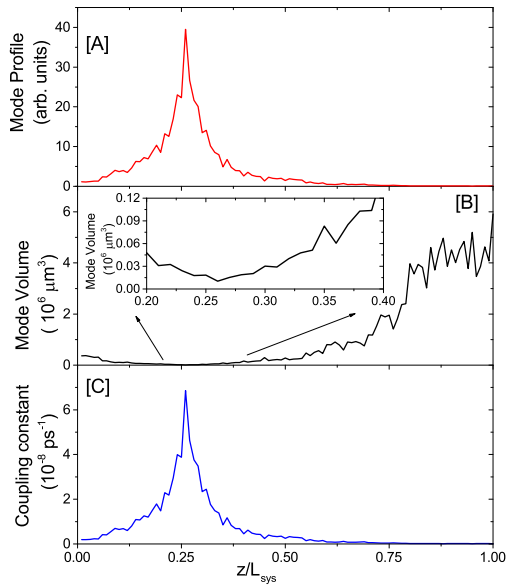}
\caption{ The mode profile ([A]) results in the spatially dependent effective mode volume ([B]), yielding a position dependent emitter-cavity coupling strength ([C]), commensurate with the mode profile.}
\end{figure}

The total outcoupled photons, or output intensity, in the $i^{th}$ mode is then computed as $n_{out,i}= \displaystyle\sum_{j}\gamma_{c,i} n_{ij}$. The cavity parameters required in the model were computed from transfer matrix method for a dielectric-air bilayer, wherein the computed structure followed the same physical parameters as in the experiments. Specifically, the dielectric layer and air layer thickness was $18~\mu$m and the number of dielectric layers was  40. The dielectric layer thickness was randomized to $\pm5$~nm to simulate intrinsic disorder, and to $\pm50$~nm to achieve extrinsic disorder. The computation also provided the longitudinal mode structure at the respective wavelength, which, with a uniform transverse structure, allowed us to calculate the effective mode volume $V_{c,ij}$. Finally, the temporal variation of the output intensity was computed using relevant parameters as follows. The $\frac{\omega_{em}}{\Delta\omega_{em}}$ was $\sim28$, assuming a 20~nm bandwidth centered at 565~nm wavelength. The $\gamma_c$ was computed from the transmission spectrum provided by the transfer matrices, while $\gamma_{rad}^{-1} \sim \gamma_2^{-1} \sim 3$~ns.. The pump rate $R_p$ took the system over threshold and was temporally varied realizing a $\sim60$~ps pulse as in the experiments.

\begin{figure}
\includegraphics[width=8.5cm]{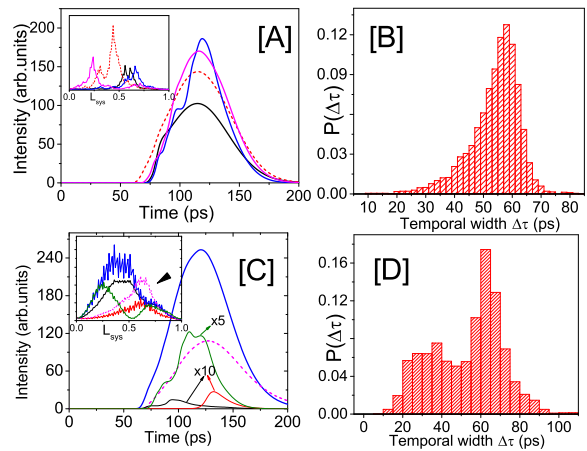}
\caption{\label{fig:figure4}  Computational results. [A] Four simultaneously excited modes in one configuration of extrinsic disorder. Inset: Spatial mode profiles in like colours. [B] Corresponding P($\Delta \tau$) over 1000 configurations, showing a left-skewed distribution. [C] Theoretically calculated results for intrinsic disorder. Five modes are shown, with their corresponding spatial profiles. [D]  Corresponding P($\Delta \tau$) reproduces the bimodal behavior.  }
\end{figure}

The results are shown in Figure 7. Fig 7[A] depicts the situation for extrinsic disorder. Multimode emission with varying profile shapes, with an occasional multipeaked character, are reproduced excellently in the computation.  The inset shows the spatial profiles of the lasing modes. Interestingly, the red mode exhibits the maximum peak intensity in the spatial distribution, but not in the temporal profile. This is the consequence of the gain redistribution in the various Anderson cavities determined by the complex interplay of mode volumes, mode profiles and quality factors. Fig 7[B] illustrates the $P(\Delta \tau)$, which peaks at $\sim 58$~ps and shows a left-skewed profile, as measured in the experiments. The computations for the intrinsic disorder are illustrated in Fig 7[C]. Five modes are excited in one configuration, which exhibit a larger variation in their peak intensities. The corresponding spatial distributions (inset) show a larger overlap, suggesting gain competition in this case. Indeed, the solid blue mode (with maximum amplitude) has a strong overlap with all but the dotted magenta mode (also marked with an arrow) which has a non-overlapping fraction. This leads to larger intensity in the (solid) blue and (dotted) magenta modes, but very small intensity in the others. The multi-peaked character is also a consequence of the mode overlap and gain competition.  Finally, the computed $P(\Delta\tau)$ (Figure 7[D]) clearly reproduces the dual-peaked behavior seen in the experiments. The long-lived modes had a $\langle \Delta \tau \rangle \sim62.5$~ps, while a new set of short-lived modes with $\langle \Delta \tau \rangle \sim 32$~ps was also sustained in the intrinsic disorder case. Thus, the computation excellently reproduced the experimental observations depicted in Fig~5.
Motivated by the agreement in the experiments and calculations, we examined the model deeper for the consequences of the modal overlap and the quality factors on the observed pulse width distribution.

\begin{figure}
\includegraphics[width=8.5cm]{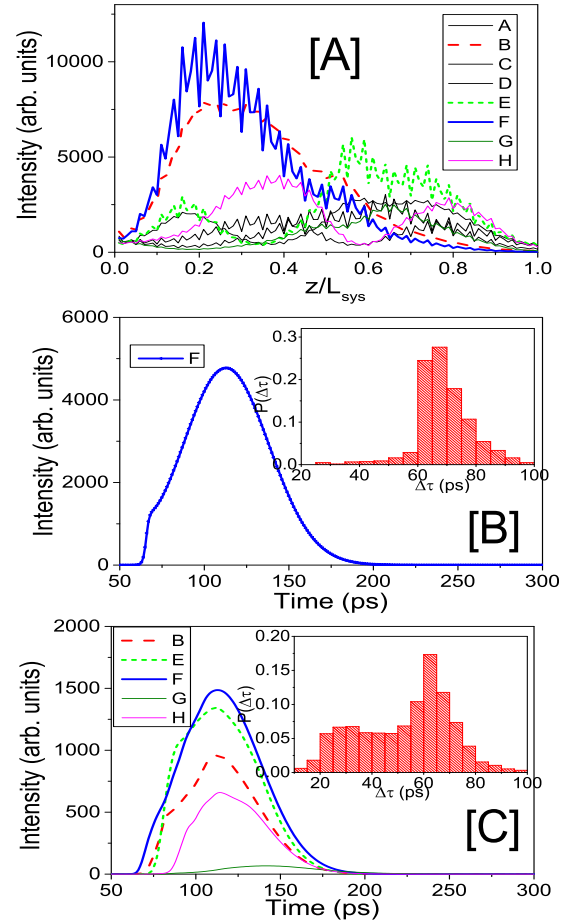}
\caption{[A] Several modes in a particular random configuration are shown. [B] The solitary lasing mode obtained when the modal spatial structure is neglected in the rate equations.  The inset shows the histogram of the temporal widths of the modes over a thousand configurations. [C] Computational result after including spatial dependence of the mode structure. As observed experimentally, multimode emission is seen. Further, the inset shows the histogram which shows a double-peak as seen in the experiments.}
\end{figure}

Figure 8 depicts the accuracy imparted by invoking the spatial structure in the rate equations.  Fig~8[A] shows several modes in a particular random configuration. As expected, one single configuration sustains multiple diversely structured modes, out of which some modes overlap substantially with each other. When the rate equations do not account for the spatial variation, all emitters are effectively posited at one point and all modes thus compete for the gain. Consequently, the strongest mode wins, as seen in Fig~8[B], where only the strongest mode  crosses the threshold. Here, the conventional rate equations are seen to predict the lasing oscillations of only the mode described by the thick blue line in [A], while the rest are all suppressed due to competition. This was a consistent observation for all configurations, which went against the experimental observation of multimode lasing. The inset in Fig~8[B] shows the histogram of the temporal widths of the modes over a thousand configurations. This histogram does not describe the experimentally observed behavior. The Fig~8[C] shows our computational result after including spatial dependence of the coupling constant. Here, while the thick-blue-line mode is stronger than others, several modes still cross the threshold depicting multimode behavior. Interestingly, the dotted-green-line mode lases stronger than the dashed-red-line mode despite the latter having a larger mode volume. This is because the dashed-red-line mode experiences competition from the thick-blue-line mode due to a near-complete overlap, whereas the dashed-green-line mode has a substantial presence in the right half of the sample not overlapping with the thick-blue-line mode. This is a very plausible and realistic behavior brought forth by the model. This facilitates the existence of several modes, agreeing with the multimode emission seen in the experiments. Importantly, the inset shows the histogram of the temporal widths, which shows the experimentally observed double-peak behavior.

\begin{figure}
\includegraphics[width=9.5cm]{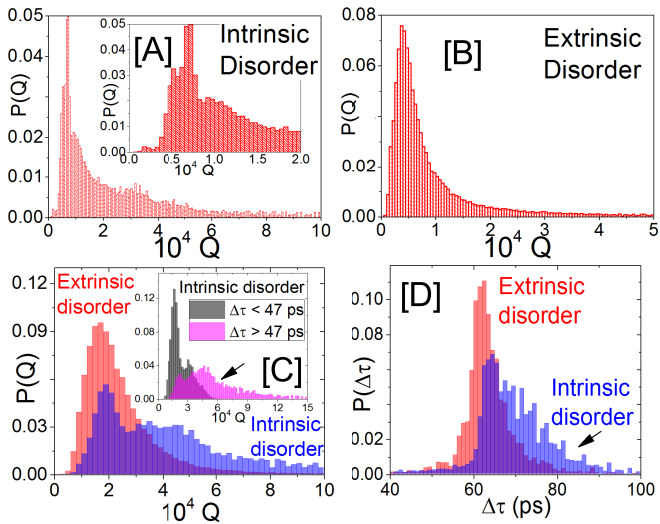}
\caption{\label{fig:figure4}  [A] Quality factor distribution $P(Q)$ of all passive cavities for intrinsic disorder. Inset: $P(Q)$ emphasized in the low-Q region. [B] $P(Q)$ for passive cavities for extrinsic disorder. [C] $P(Q)$ for lasing cavities for intrinsic disorder (blue histogram) and extrinsic disorder (red histogram). Inset: $P(Q)$ for the low $\Delta \tau$ (black histogram) and high $\Delta \tau$ (magenta histogram) bunch seen in Fig 4[D]. [D] Computed P($\Delta \tau$) when the spatial mode structure is turned off.}
\end{figure}

Fig~9[A] illustrates the quality factor distributions $P(Q)$ of the passive cavities realized in the structure with intrinsic disorder. The distribution has a very long tail, with the average Q $2.4 \times 10^4$. The $P(Q)$ exhibits a structure in the low-Q region, which is emphasized in the inset. [B] exhibits $P(Q)$ for the extrinsic disorder, where the distribution is narrower and smoother, with an average Q of $1.7 \times 10^4$. Such reduction in quality factor with increasing disorder was also observed in disordered photonic crystal waveguides\cite{liu14}. Fig 9[C] shows the $P(Q)$ for lasing cavities, i.e., only the cavities that cross the lasing threshold. Indeed, in the intrinsic disorder (blue histogram), the crystal turns out to have a multibunched distribution, a fact which sets the system apart from the PC waveguides where no such effect was observed at least at the reported intrinsic disorder\cite{liu14}. An increase in the disorder strength defeats this nature and the distribution becomes single-peaked (Fig~9[C], red histogram). However, it should be noticed that a significant overlap exists between the two distributions at the low-Q region ($Q \sim 20000$). In contrast, the $P(\Delta\tau)$ (Fig~7[B] and [D]) overlap in the region of large $\Delta\tau$. We further categorised the long-lived and short-lived modes in intrinsic disorder separately, by tentatively demarcating them at $\Delta\tau = 47$~ps. The inset in Fig~9[C] shows the $P(Q)$ for the intrinsic disorder for modes with $\Delta\tau < 47$~ps (black histogram) and other for $\Delta\tau > 47$~ps (magenta histogram). The substantial overlap in these two histograms also implies that the $\Delta\tau$ is not trivially determined by the Q-factor. The random mode structure of the Anderson cavity plays a pivotal role in determining the $P(\Delta\tau)$. This fact is evidenced by the distributions in Fig~9[D]. Here, we compute the $\Delta\tau$ for conventional cavities (by ignoring the random mode structure) with the same $P(Q)$ as in Fig~9[C]. Here, the two systems are seen to behave comparably. Thus, the mode structure, mode overlap and consequent gain competition play a significant role in the behavior of $P(\Delta\tau)$. Overall, these computations essentially emphasize the complexity in the temporal dynamics of the system, wherein no single cavity parameter  such as Q-factor, mode volume, mode profile etc can trivially determine the behavior, but there is a collective contribution by all the parameters that manifest the temporal fluctuations.

In summary, we have reported the temporal complexity in the pulsed emission of Anderson localized lasers. We have presented the direct measurements of the exponentially-decaying spatial profiles and temporal width distributions of the emission pulses, and have reported a novel disorder-dependence. Under extrinsic disorder, the pulsewidth distribution exhibited a single-peaked left-skewed behavior. For intrinsic disorder, a bimodal distribution, with an additional short-timescale peak, was seen. Our computations based on modified coupled emitter-cavity rate equations are in excellent agreement with the experiments, and reveal the pivotal role played by the mode structure and consequent gain competition. These results are likely to open a new arena in the investigations of Anderson localization with gain.

\section{Acknowledgements}
We acknowledge helpful discussions with Prof Martin Kamp, University of Wuerzburg, and Anjani Kumar Tiwari. SM gratefully acknowledges financial support for the project `XIIP243: Optics in nanostructures', funded by the DAE, Government of India, and the Swarnajayanti Fellowship from the DST, Government of India. SM dedicates this work to Late Prof Narendra Kumar, who pioneered the research on the synergy of optical amplification and Anderson localization\cite{pradhan94}.

\end{document}